\documentclass[twocolumn,showpacs,amsmath,amssymb,prl]{revtex4}

\pdfoutput=1

\usepackage{graphicx}
\usepackage{dcolumn}
\usepackage{bm}
\usepackage{epsf}

\newcommand{\bra}[1]{\langle #1|}
\newcommand{\ket}[1]{|#1\rangle}
\newcommand{\braket}[2]{\left\langle #1|#2\right\rangle}
\newcommand{\tr}[1]{\text{tr}\left\{#1\right\}}
\newcommand{\la}{\left\langle}
\newcommand{\ra}{\right\rangle}

\newcommand{\bla}{bla\\bla\\bla\\bla\\bla}
\newcommand{\PRA}{Phys. Rev. A }
\newcommand{\PRD}{Phys. Rev. D }
\newcommand{\PRE}{Phys. Rev. E }
\newcommand{\PRL}{Phys. Rev. Lett. }
\newcommand{\EPL}{Europhys. Lett. }

\begin{document}

\title{Generalized Clausius inequality for nonequilibrium quantum processes}
\author{Sebastian Deffner}
\author{Eric Lutz}
\affiliation{Department of Physics, University of Augsburg, D-86135 Augsburg, Germany}
\date{\today}

\begin{abstract}
We show that the nonequilibrium entropy production for a driven quantum system is larger than the Bures length, the geometric distance between its actual state and the corresponding equilibrium state. This universal lower bound generalizes the Clausius inequality to arbitrary nonequilibrium processes beyond linear response. We further derive a fundamental upper bound for the quantum entropy production rate and discuss its connection to the \mbox{Bremermann-Bekenstein bound}. 
\end{abstract}
\maketitle

All real macroscopic processes are irreversible.  In thermodynamics,  irreversibility is quantified by means of the entropy $S$: For any state transformation, the variation of entropy is written as $\Delta S = \Delta S_\text{re} + \Delta S_\text{ir}$, where  $\Delta S_\text{re}= Q/T$ is the entropy change associated with reversible (equilibrium) processes \cite{gro84}. The Clausius inequality, $\Delta S_\text{ir}\geq0$, on the other hand, provides a fundamental characterization of irreversible (nonequilibrium) phenomena by specifying a lower bound for the irreversible entropy change;  this lower bound (zero) is trivially  independent  of how far from equilibrium a process operates. In many cases of interest, however, having  a sharper, transformation dependent lower bound is essential. A case in point is the  optimization of the performance of  real thermodynamic processes that occur in finite time \cite{and84,sal98}. For classical, near-equilibrium transformations, such a lower bound has been derived using a geometric approach to thermodynamics \cite{rup95}: The infinitesimal irreversible entropy production is   given by the Riemannian distance between initial (equilibrium) and final (nonequilibrium) states, $d S_\text{ir}\gtrsim d \ell^2/2 $ \cite{sal83,sal84,nul85,cro07}; this expression  is obtained  by a second-order expansion around equilibrium and is therefore restricted to the linear response regime. The thermodynamic length $\ell$  measures the number of distinguishable states between  initial and final probability distributions, $p_0$ and $p_\tau$,  and  is identical to Wootters' statistical distance  between wave vectors (pure states)  in Hilbert space \cite{woo81}; it is explicitly given by 
\begin{equation}
\label{q01}
\ell\left( p_0,p_\tau\right) =\arccos{\left( \int dx\, \sqrt{p_0\left( x\right) \, p_\tau\left( x\right) } \right) }\ ,
\end{equation}
and  hence measures the angle  in state space separating  the two probability distributions $p_0$ and $p_\tau$.

In this paper, we extend the above results to quantum, far from equilibrium transformations. Specifically, we generalize the familiar Clausius inequality by  deriving a universal lower bound for the irreversible entropy production valid for arbitrary quantum processes, using tools from quantum information theory. We moreover show that, in contrast to classical nonequilibrium physics, there exists a maximum  quantum entropy production rate. 
Our work is motivated by recent experiments on driven cold-atom gases that  for the first time allow the  direct investigation of the nonequilibrium dynamics of isolated many-particle quantum systems beyond the near-equilibrium  linear response regime \cite{kin06,hof07}. A precise characterization of the quantum nonequilibrium  entropy production in this unchartered domain appears therefore necessary. In the following, we start from a recently derived formula for the total work done on a closed quantum system \cite{tal07} to obtain a microscopic expression for the irreversible entropy production, valid  far from equilibrium. We show that the latter is bounded from below by the Bures length \cite{kak48,bur69}, a quantum generalization for mixed states of  Wootters' statistical distance. The Bures length is closely related to the quantum  fidelity, a central measure of quantum information theory \cite{nie00}. We further demonstrate that the quantum entropy production rate $\sigma$   is bounded from above by a quantity that also depends on the Bures length. This fundamental limit  on the entropy variation rate is of purely quantum origin and is connected to  the energy-time uncertainty relation. Remarkably, we show that the maximum rate reduces  to the Bremermann-Bekenstein bound on \mbox{information flow \cite{bek81}}.

\paragraph{Quantum nonequilibrium entropy production.} Let us consider a closed quantum system, initially in a thermal state, whose Hamiltonian $H_t$ is driven by an external time-dependent parameter during time $\tau$. When this parameter is changed in a slow, quasistatic way, the system remains in  equilibrium and the process is reversible. By contrast, for a fast change, for instance a parameter quench, the system is driven in a nonequilibrium state. The nonequilibrium entropy variation associated with such a transformation may be defined as \cite{jar97},
\begin{equation}
\label{q02}
\Delta S_\text{ir}= \beta\la W_\text{ir} \ra \ ,
 \end{equation}
where $\la W_\text{ir} \ra=\la W \ra-\Delta F$ is the difference between the total work $\la W \ra$ done on the system during time $\tau$ and the free energy difference  $\Delta F$ (the reversible work); as usual $\beta = 1/(kT)$ denotes the inverse temperature. The irreversible work $\la W_\text{ir}\ra$ vanishes for a reversible process and is  defined even if the final state of the system is arbitrarily far from equilibrium.
 The probability distribution of quantum work  is  given by the difference $E_{m}^{\tau }-E_{n}^{0}$ of  final and initial system energy eigenvalues, averaged over all initial states (thermal distribution $p_n^0 = \exp(-\beta E_n^0)/Z_0)$ and final states (transition probabilities $p_{m,n}^{\tau}$) \cite{tal07},
\begin{equation}
\label{q03}
{\cal P}\left( W\right) =\sum_{m,n}\,\delta\left( W-\left( E_{m}^{\tau }-E_{n}^{0}\right) \right) \,p_{m,n}^{\tau}\, p^0_{n} \ .
\end{equation}
An experimental scheme to  measure  the distribution ${\cal P}\left( W\right) $ in a  modulated cold ion trap has  been proposed  in Ref.~\cite{hub08}. According to Eq.~\eqref{q03}, the mean  work is simply
$\la W \ra=\sum_{m,n}\,( E_{m}^{\tau }-E_{n}^{0} ) \,p_{m,n}^{\tau}\, p^0_{n}$.
By introducing the equilibrium density operator at the final time $\tau$, \mbox{$\rho_\tau^\text{eq}=\exp{\left( -\beta H_\tau \right) }/Z_\tau$} with eigenvalues $ p_m^\tau$, we can write 
\begin{eqnarray}
\label{q04}
\la W \ra&=&1/\beta\,\sum\limits_n\, p^0_n \,\ln{ p^0_n }-1/\beta\,\sum\limits_{m,n}\,p_n^0\, p_{m,n}^{\tau}\,\ln{ p_m^\tau }\nonumber\\
&-&1/\beta\,\ln{\left(Z_\tau/Z_0\right) }\ .
\end{eqnarray}
The last term on the right-hand side is equal to $\Delta F$, while the first two are $(1/\beta)$ times the quantum  
Kullback-Leibler divergence $S(\rho_\tau||\rho_\tau^\text{eq})$, or quantum relative entropy \cite{ume62}, between the actual density operator of the system $\rho_\tau$ at time $\tau$ and the corresponding equilibrium density operator $\rho_\tau^{\text{eq}}$. Using Eq.~\eqref{q02}, we therefore obtain,
\begin{equation}
\label{q05}
\Delta S_\text{ir}=S\left( \rho_\tau||\rho_\tau^\text{eq} \right) =\tr{\rho_\tau\ln{\rho_\tau}-\rho_\tau\ln{\rho_\tau^{\text{eq}}}}\,.
\end{equation}
This is an exact expression for the nonequilibrium entropy production and a quantum generalization of recent results presented in Refs.~\cite{kaw07,vai09}. We note, however, that the relative entropy is not a true metric, as it is not symmetric and does not satisfy the triangle inequality; it can therefore not be used as a proper quantum distance  \cite{yeu02}. Furthermore, Eq.~\eqref{q05} is in general difficult to determine explicitly. We next derive a lower bound for the quantum  entropy production which we express in terms of the fidelity, one of the most commonly used and well-studied measures in quantum information theory \cite{nie00}.

\paragraph{Generalized quantum Clausius inequality.} Inequalities are  essential tools of classical and quantum information theory \cite{yeu02}; they allow to express 'impossibilities', things that cannot happen, and relate hitherto unconnected  quantities. An elementary example is Klein's inequality,  $S( \rho_1||\rho_2) \geq 0$, which asserts the non-negativity of  the quantum relative entropy \cite{nie00}. Combined with  Eq.~\eqref{q05}, it immediately leads to the usual Clausius inequality. We shall establish a generalized Clausius inequality by proving that the irreversible entropy variation is always larger than the Bures length \cite{kak48,bur69}. The Bures metric  formally quantifies the infinitesimal distance between two density operators as ${\cal L}^2(\rho+\delta\rho,\rho)=\tr{\delta\rho\, G} /2$, where $G$ obeys $\rho\,G+G\,\rho=\delta\rho$.  Distances should be physically motivated and  to some degree unique \cite{rup95}. Wootters' statistical distance,  being equal to the angle in Hilbert space, is the only Riemannian metric (up to a constant factor) which is invariant under {\it all} unitary transformations \cite{woo81}; it is hence a natural metric on the  space of pure states. The Bures metric, on the other hand, is the generalization of Wootters' metric to mixed states \cite{bra94}; in this sense it represents a natural, unitarily invariant  Riemannian metric on the space of impure density matrices \cite{ben06}.  For  any two density operators the finite Bures length ${\cal L}$  is given by
\begin{equation}
\label{q07}
{\cal L}\left(\rho_1,\rho_2\right)=\arccos{\left(\sqrt{F\left(\rho_1,\rho_2\right)}\right)}\,,
\end{equation}
where the fidelity $F$ is defined for an arbitrary pair of mixed quantum states as \cite{uhl76,joz94},
\begin{equation}
\label{q08}
F\left( \rho_1,\rho_2\right)=\left[\tr{\sqrt{\sqrt{\rho_1}\,\rho_2\, \sqrt{\rho_1}}} \right]^2\,.
\end{equation} 
The fidelity is a symmetric, non-negative and unitarily invariant function, which is equal to one only when the two states are identical. For pure quantum states, $\rho_i=\ket{\psi_i}\bra{\psi_i}$, the fidelity reduces to their overlap, $F( \rho_1,\rho_2)=\tr{ \rho_1 \rho_2} = |\braket{\psi_1}{\psi_2} |^2$.
It has recently been shown that if $d(\rho_1,\rho_2)$ is a unitarily invariant  norm, then the quantum relative entropy satisfies (Ref.~\cite{aud05}, Th. 4), 
\begin{equation}
\label{q09}
S(\rho_1||\rho_2)\geq 2\,\frac{d^2\left(\rho_1,\rho_2\right)}{d^2\left(e^{1,1},e^{2,2}\right)}\,,
\end{equation}
where $e^{i,j}$ is the matrix with $i, j$ element equal to 1 and all other elements 0. Noting that  ${\cal L}(e^{1,1},e^{2,2}) = \pi/2$, since the two matrices are orthogonal ($F(e^{1,1},e^{2,2}) =0$), we obtain 
the generalized Clausius inequality,
\begin{equation}
\label{q10}
\Delta S_\text{ir} \geq\frac{8}{\pi^2} \,{\cal L}^2\left(\rho_\tau,\rho_\tau^\text{eq}\right)\,.
\end{equation}
The  quantum entropy production $\Delta S_\text{ir}$ is hence bounded from below by the geometric distance  between the actual density operator $\rho_\tau$  at the end of the process and the corresponding equilibrium operator $\rho_\tau^\text{eq}$, as measured by the Bures length; the latter defines a quantum generalization of the concept of thermodynamic length \cite{sal83,sal84,nul85,cro07} and provides a natural scale to compare $\Delta S_\text{ir}$ with.  In other words, inequality  \eqref{q10} quantifies in a precise way the intuitive notion that the irreversible entropy production is larger when a system is driven farther away from equilibrium.  Expression  \eqref{q09} shows that $\Delta S_\text{ir}$ is bounded by many distances, however, only the Bures length has a simple physical interpretation. Equation \eqref{q10} is valid for arbitrary quantum processes, including far from equilibrium final states. For infinitesimally close diagonal states, we have $S(\rho||\rho+d\rho) \simeq 2 {\cal L}^2(  \rho||\rho+d\rho) \simeq d \ell^2(\rho||\rho+d\rho)/2$. In the limit of classical, quasi-equilibrium transformations, Eq.~\eqref{q10} thus reduces to $dS_\text{ir} \geq 2/\pi^2\, d\ell^2$.

\begin{figure}
\centering
\includegraphics[width=0.46\textwidth]{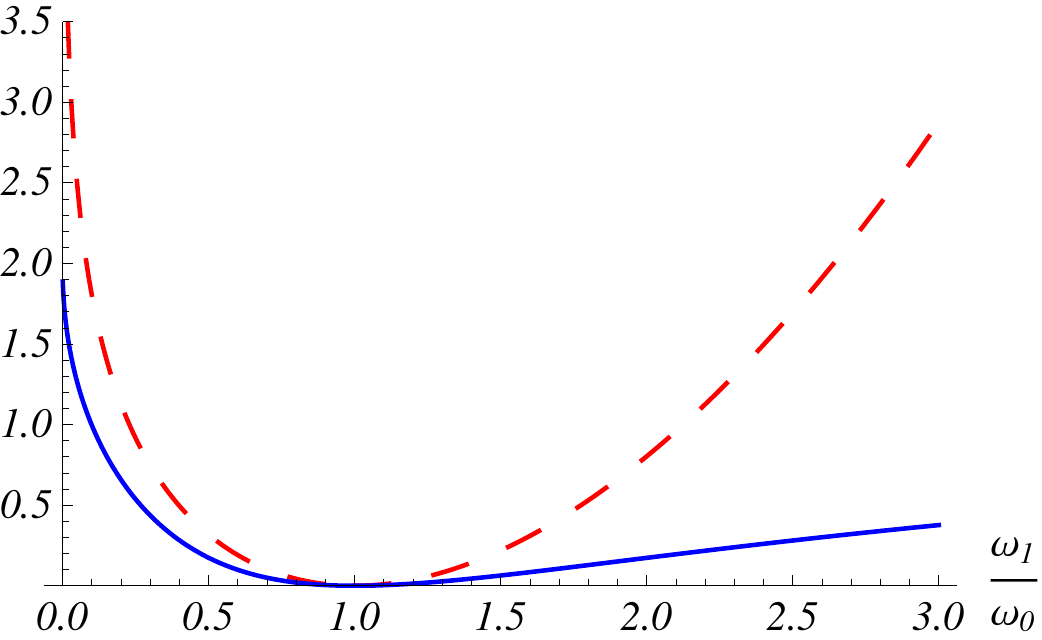}
\caption{(Color online) Illustration of the generalized Clausius inequality \eqref{q10} for the quantum harmonic oscillator \eqref{q11} with linearly varying frequency.  Entropy production $\Delta S_\text{ir}$ (red, dashed) and lower bound $ 8/\pi^2\,{\cal L}^2\left(\rho_\tau,\rho_\tau^\text{eq}\right)$ (blue)  as a function of the final frequency $\omega_1$  for $\hbar=\beta=\tau=1$ and $\omega_0=0.1$.} 
\label{fig1}
\end{figure}

A simple illustration of the generalized Clausius inequality \eqref{q10} is provided by  a time-dependent quantum harmonic oscillator, initially at thermal equilibrium,
\begin{equation}
\label{q11}
H_t=\frac{p^2}{2m}+\frac{1}{2}m \omega_t^2\,x^2 \ ,
\end{equation}
and whose angular frequency is varied from $\omega_0$ to $\omega_1$ according to $\omega_t^2=\omega_0^2+ (\omega_1^2-\omega_0^2)\,t/\tau$ (cf. Fig.~\ref{fig1}) \cite{def08}. This quantum system is exactly solvable and both the entropy production and the Bures length can be evaluated explicitly \cite{def10}.
The distance from equilibrium is here controlled by the ratio $\omega_1/\omega_0$ between  final and initial  frequencies. We observe that the relation $dS_\text{ir} \simeq d \ell^2/2$ is valid close to equilibrium, $\omega_1/\omega_0\simeq 1$, whereas the generalized Clausius inequality \eqref{q10} holds  for all values of $\omega_1$. It is worth noting that for very large $\omega_1$, initial and final state become maximally distinguishable, i.e. orthogonal, and the Bures length \eqref{q07} approaches $\pi/2$. For the parameters of   Fig.~\ref{fig1}, this very far from equilibrium regime is reached when the energy of the oscillator is increased by more than a factor of twenty. A quantitative estimate of the energy change associated with a given irreversible entropy production will be given below.

\paragraph{Quantum entropy production rate.}
Nonequilibrium irreversible phenomena are not only characterized by the irreversible entropy change, but also by the rate of entropy production $\sigma=\Delta S_\text{ir}/\tau$. The entropy rate $\sigma$ is a central quantity that is associated with  the speed of evolution of a nonequilibrium process \cite{gro84}. In quantum mechanics, the energy of a system imposes a fundamental constraint on its unitary time evolution, as captured  for instance by the time-energy uncertainty relation $t\geq \hbar/\Delta E$: the minimal time it takes for a quantum system to evolve to an orthogonal state is always larger than the inverse of its initial energy spread $\Delta E$. A more accurate expression for the 'quantum speed limit' has been derived in Ref.~\cite{gio03} for time-independent Hamiltonians. Consider a system initially in state $\rho_0$ with mean energy $E_0 = \la H_0 \ra$ and energy spread $\Delta E_0 =(\la H_0^2\ra-\la H_0\ra^2)^{1/2}$. The minimum time required for the evolution to an arbitrary  state $\rho_\tau$ is then given by
\begin{equation}
\begin{split}
\label{q12}
\tau_\text{min}\simeq \max{\left\{\frac{2\hbar\,{\cal L}^2\left(\rho_\tau,\rho_0\right)}{\pi \,E_0},\, \frac{\hbar\,{\cal L}\left(\rho_\tau,\rho_0\right)}{\Delta E_0}\right \}}\ ,
\end{split}
\end{equation}
where ${\cal L}(\rho_\tau,\rho_0)$ is the Bures length \eqref{q07} between $\rho_\tau$ and $\rho_0$. The quantum speed limit time is thus entirely determined by the initial energy (mean or variance)  and the geometric distance between initial and final states. We will now show that Eq.~\eqref{q12} sets an upper bound to the entropy production rate. We begin by rewriting Eq.~\eqref{q01} in the form $\Delta S_\text{ir}=\beta | \la H_\tau\ra- \la H_0\ra - F_\tau+ F_0|$, since $\Delta S_\text{ir}\geq0$, and focus  on the limit of large excitations $\la H_\tau\ra\gg\la H_0\ra$, which is achieved for long enough driving times. By applying the triangle inequality and noting that $\Delta F=F_\tau- F_0 \leq \la H_\tau\ra- \la H_0\ra $, we obtain
\begin{eqnarray} 
\label{q13}
\Delta S_\text{ir}&\leq&\beta(\la H_\tau\ra+ \la H_0\ra + \Delta F) \nonumber \\
&\leq& 2 \beta(\la H_\tau\ra+ \la H_0\ra )\simeq 2 \beta \la H_\tau\ra \ ,
\end{eqnarray}
By combining Eqs.~\eqref{q12} and \eqref{q13}, we find that the maximum entropy production rate $\sigma_\text{max}=\Delta S_\text{ir,max}/\tau_\text{min}$ is
\begin{equation}
\label{q14}
\sigma_\text{max}\simeq 2\beta\la H_\tau\ra\min{\left\{\frac{ \pi\, E_0 }{2\hbar\,{\cal L}^2\left(\rho_\tau,\rho_0\right)}, \frac{ \Delta E_0}{\hbar \,{\cal L}\left(\rho_\tau,\rho_0\right)}  \right\}}
\end{equation}
The above equation expresses the inherent quantum-mechanical limit to the nonequilibrium entropy production rate. Equation \eqref{q14} simplifies in the limit of far from equilibrium transformations when initial and final states become orthogonal, ${\cal L}\left(\rho_\tau,\rho_0\right)\simeq\pi/2$, and in the limit of high temperatures, $E_0\simeq \Delta E_0 \simeq 1/\beta$, to the Bremermann-Bekenstein bound  \cite{bek81},
\begin{equation}
\label{q15}
\sigma \leq\frac{4}{\hbar\pi}\,\la H_\tau\ra\ .
\end{equation} 
The Bremermann-Bekenstein bound on  information flow gives the maximum quantum communication rate (capacity) that is possible through a noiseless single channel with signals of finite duration. We stress that the present derivation is  solely based on the thermodynamic definition  of the entropy production \eqref{q02} and does not make  any reference to information entropy or  channels.; it is thus free of the caveats of the original derivations, such as the use  of the classical  Shannon formula for the capacity, or the periodic boundary condition approximation \cite{bek90}.
\begin{figure}
\centering
\includegraphics[width=0.47\textwidth]{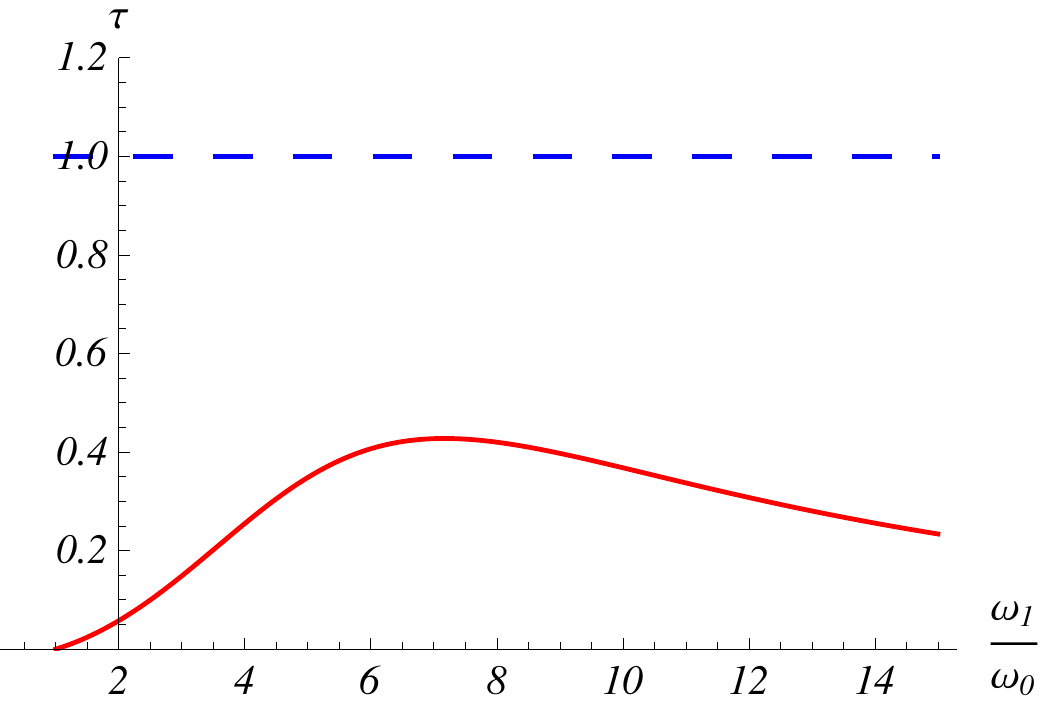}
\caption{(Color online)  Quantum speed limit time $\tau_\text{min}$, Eq.~\eqref{q14}, (red) and actual driving time $\tau$ (blue, dashed) 
for the linearly parameterized  quantum harmonic oscillator \eqref{q11} with $\hbar=\tau=1$, $1/\beta=0$ and $\omega_0=0.1$.} 
\label{fig2}
\end{figure}

The quantum speed limit time  \eqref{q12}, and hence the entropy production rate \eqref{q14}, only hold for slowly driven systems (i.e. quasi-equilibrium processes), as Eq.~\eqref{q12} assumes a time-independent   Hamiltonian. For arbitrary nonequilibrium transformations, Eq.~\eqref{q12} can be extended  using the geometric approach of Ref.~\cite{jon10}.  By evaluating the time derivative of  the angle between initial and final state, ${\cal L}\left(\rho_\tau,\rho_0\right)$, we find that the exact quantum speed limit for time-dependent Hamiltonians is 
\begin{equation}
\label{q16}
\tau_\text{min}=\max{\left\{\frac{\hbar\,{\cal L}\left(\rho_\tau,\rho_0\right)}{E_\tau},\frac{\hbar\,{\cal L}\left(\rho_\tau,\rho_0\right)}{\Delta E_\tau}\right\}}\,.
\end{equation}
Contrary to Eq.~\eqref{q12}, the minimum time  is here determined by   the time averaged mean energy and energy variance, 
$E_\tau=(1/\tau)\int_0^\tau dt \la H_t\ra$ and  $\Delta E_\tau=(1/\tau)\int_0^\tau dt \,(\la H_t^2\ra -\la H_t \ra^2)^{1/2}$, and not by their initial values \cite{rem}. As a consequence,  the quantum speed limit time for driven systems can be smaller than for undriven systems  when $E_\tau> E_0$ ($\Delta E_\tau> \Delta E_0$). This is for instance the case at zero temperature: According to Eq.~\eqref{q12}, a quantum system never leaves an initial pure  state (infinite $\tau_\text{min}$) in the absence of driving, while  Eq.~\eqref{q14} predicts a finite $\tau_\text{min}$ for a driven Hamiltonian. Figure \ref{fig2}  shows that for the time-dependent  oscillator \eqref{q11} at zero temperature, the actual driving time $\tau$ can approach the absolute minimum evolution time  $\tau_\text{min}$ within a factor of two, for a simple linear change of its angular frequency. 

The general expression for the maximum entropy production rate $\sigma$ for nonequilibrium quantum processes can eventually be obtained by combining Eqs.~\eqref{q12} and \eqref{q16}; it is to be regarded as the extension of the Bremermann-Bekenstein bound \eqref{q15} to arbitrary distances between initial and final states, arbitrary initial temperature and arbitrary transformation speed. Written in the form
\begin{equation}
\label{q17}
\frac{\la H_t\ra}{ \Delta S_\text{ir}} \geq \frac{2\beta}{\tau}\min{\left\{\frac{E_\tau}{\hbar\,{\cal L}\left(\rho_\tau,\rho_0\right)},\frac{\Delta E_\tau}{\hbar\,{\cal L}\left(\rho_\tau,\rho_0\right)}\right\}}\ ,
\end{equation}
it provides an estimate for the minimum energy change occurring with a given entropy variation in a time $\tau$.

\paragraph{Conclusion.}
We have developed a generic  geometric characterization of far from equilibrium quantum processes based on the distinguishability metric on the space of quantum states. We have first obtained a generalized, more precise form of the Clausius inequality by deriving a lower bound for the irreversible entropy production given by the Bures length between the  nonequilibrium density operator of the system and the corresponding equilibrium operator. We have further established the existence of an upper bound for the entropy production rate by employing the notion of quantum speed limit time which is itself a function of the Bures length. The latter is an extension of the Bremermann-Bekenstein bound to arbitrary nonequilibrium quantum processes.

This work was supported by  the Emmy Noether Program of the DFG (contract No LU1382/1-1) and the cluster of excellence Nanosystems Initiative Munich (NIM).


\begin{thebibliography}{99}
\bibitem{rem} Note that the exact $\tau_\text{min}$ \eqref{q16} does not reduce exactly to the estimated $\tau_\text{min}$ \eqref{q12} in the limit of slow driving.
\end{thebibliography}
\end{document}